\RequirePackage{fix-cm}
\documentclass[smallcondensed]{svjour3}     
\smartqed  
\usepackage{graphicx}
\usepackage[utf8]{inputenc}
\usepackage{amssymb}
\usepackage{amsmath}
\usepackage{amsfonts}
\usepackage{color}
\usepackage{graphicx}
\usepackage{hyperref}

\usepackage{blindtext}
\usepackage{array}
\usepackage{tabularx}
\graphicspath{{./Images/}}
\usepackage{hyperref}
\usepackage{caption}
\usepackage{subcaption}
\usepackage{float}
\graphicspath{{./images/}}
\usepackage{rotating}

\begin{document}

\title{Deep Learning for Low-Dose CT Denoising Using Perceptual Loss and Edge Detection Layer
}

\titlerunning{Deep Learning for Low-Dose CT Denoising}        
\author{Maryam~Gholizadeh-Ansari \and
        Javad~Alirezaie \and
        ~Paul~Babyn
}
\institute{	
Maryam Gholizadeh-Ansari \and Javad Alirezaie \at
Department of Electrical and Computer Engineering, Ryerson University, Toronto, ON, M5B2K3, Canada \\
Tel.: +1-416-979-5000\\
Fax: +1-416- 979-5280\\
\email{ansari@ryerson.ca, javad@ryerson.ca}
\and
Paul Babyn \at
Department of Medical Imaging, University of Saskatoon Health Region, Royal University Hospital, Saskatoon, SK, S7N0W8, Canada \\
\email{paul.babyn@saskatoonhealthregion.ca}
}
\date{Received: date / Accepted: date}

\maketitle

\begin{abstract}
Low-dose CT denoising is a challenging task that has been studied by many researchers. Some studies have used deep neural networks to improve the quality of low-dose CT images and achieved fruitful results.  In this paper, we propose a deep neural network that uses dilated convolutions with different dilation rates instead of standard convolution helping to capture more contextual information in fewer layers. Also, we have employed residual learning by creating shortcut connections to transmit image information from the early layers to later ones. To further improve the performance of the network, we have introduced a non-trainable edge detection layer that extracts edges in horizontal, vertical, and diagonal directions. Finally, we demonstrate that optimizing the network by a combination of mean-square error loss and perceptual loss preserves many structural details in the CT image. This objective function do not suffer from over smoothing and blurring effects caused by per-pixel loss and grid-like artifacts resulting from perceptual loss. The experiments show that each modification to the network improves the outcome while only minimally changing the complexity of the network.

\end{abstract}

\section{Introduction}
Computed tomography (CT) is an accurate and non-invasive method to detect internal abnormalities of the body such as tumors, bone fractures, and vascular disease. It has been widely used by clinicians to diagnose and monitor conditions such as cancer, lung disease, and abnormalities of the internal organs.  

As CT images are produced by transmitting X-ray beams through the body, there has been growing concern about the risk of CT radiation. The amount of exposure during one session of CT scan is much higher than a conventional X-ray. For example, the radiation that a patient receives in a chest X-ray radiography is equal to 10 days of background radiation \cite{radiation-dose}. Background radiation is the amount of radiation that a person gets from cosmic and natural resources in daily life. During a chest CT scan, the radiation exposure is equal to two years of background radiation \cite{radiation-dose}. Therefore, the radiation risk is much higher in computed tomography especially for those who require multiple CT scans. While radiation affects all age groups, children are more vulnerable than adults because of their developing body and the longer lifespan. Research has found that children who have cumulative doses from multiple head scans have increased risk (up to three times increased risk of diseases such as leukemia and brain tumors \cite{radiation-risks}. 

Considering the advantages of CT scans for diagnosis, it is critical to find a solution to minimize radiations. One approach to decreasing the radiation risk is to use lower levels of X-ray current; however, these CT images have increased noise and may not be as diagnostic. 

In recent years, many types of research have been conducted to enhance the quality of the reconstructed CT images. Researchers have followed three paths to remove noise from low-dose CT images: processing the raw data obtained from sinograms (projection space denoising), iterative reconstruction methods, and processing reconstructed CT image (image space denoising) \cite{CT-denoising-methods}. 

In projection space denoising, the noise removal algorithm is applied to the CT sinogram data obtained from low-dose X-ray beams.  Sinogram data, also called projection or raw data, is a 2-D signal that represents the sum of the attenuation coefficients for a beam passing through the body. The noise distribution of low-dose CT image in the projection space can be well-characterized \cite{poisson-noise1,poisson-noise2} which makes the noise removal task simple. Some researchers have applied traditional noise removal techniques on this data including bilateral filtering before image reconstruction \cite{bilateral,sinogram-denoising}. These methods incorporate system physics and photon statistics to reduce both noise and artifacts. However, it makes the algorithm, vendor dependent. These methods also need access to sinogram data which is generally not available for many commercial CT scanners. Finally, these techniques should be implemented on the scanner reconstruction system and increases the cost of denoising \cite{CT-denoising-methods}.  

Iterative reconstruction methods are other means to improve the quality of low-dose CT images \cite{MBIR,SAFIRE1}. In these methods, the data is transformed to the image domain and projection space multiple times to optimize the objective function. In the first step, a CT image is reconstructed using the projection data and then it is transformed back to the projection space. In each iteration, the generated projection data from the reconstructed CT image is compared with the actual data from the scanner to get improved. The process stops when the convergence criteria are met. These methods may take into account system model geometry, photon counting statistics, as well as x-ray beam spectrum and they usually outperform projection space denoising methods. Iterative techniques are capable of removing artifacts and providing good spatial resolution. However, similar to the previous group, they need access to the projection data, are vendor dependent and should be implemented on the reconstruction system of the scanner. Moreover, the process is slow, and the computational cost of multiple iterations is high \cite{CT-denoising-methods}. 

Unlike to the previous methods, image space denoising algorithms do not require the projection data. They work directly on the reconstructed CT images and are generally fast, independent of the scanner vendor and can be easily integrated with the workflow. Many of the proposed algorithms in this category are adopted from natural image processing. KSVD \cite{ksvd} is a dictionary learning algorithm based on sparse representation and dictionary learning. It is used for tasks such as image denoising, feature extraction, and image compression. In some studies, KSVD is employed to improve the quality of low-dose CT scans \cite{abdomen-ldct-ksvd,abhari-ldct-ksvd}. Non-local means \cite{non-local-means} is another algorithm initially proposed for image denoising that has also been used for low-dose CT image enhancement \cite{ldct-non-local-means}. The method calculates a weighted mean of the pixels in the image based on their similarity to the target pixel. The state of the art block matching 3D (BM3D) \cite{bm3d} is also proposed for dealing with natural image noise.  It is similar to the non-local means but works in a transform domain like wavelet or discrete Fourier transform. The first step of BM3D is to group patches of the image that have similar local structure and then stack them and form a 3-dimensional array. After transforming the data, a filter is applied to remove the noise. This method has been followed in some studies to perform low-dose CT noise removal \cite{ldct-bm3d-1,ldct-bm3d-2}.  

Recently, many advances have been made in the image processing field using deep learning (DL). The high computational capacity of GPUs in combination with techniques such as batch normalization \cite{batch-normalization} and residual learning \cite{residual-learning} have made training deep networks possible. Some of those proposed networks have outperformed traditional methods in challenging tasks such as image segmentation, image recognition, and image enhancement. Medical imaging has also benefitted from this advancement. One of the first networks to reduce the noise of the low-dose CT image was proposed by Chen et al. \cite{chen}. It was inspired from a network designed for image super-resolution with three convolutional layers \cite{sr-cnn}. Convolutional auto-encoders have been used in \cite{ldct-encoder-1}and \cite{ldct-encoder-2} while the later  also takes advantage of residual learning. All of the mentioned networks offer an end to end solution for low-dose noise removal. They receive a low-dose CT image as an input and predict the normal-dose CT image as the output. However, Kang et al. firstly find the wavelet coefficients for low-dose and normal-dose CT images \cite{ldct-wavelet}. Then these wavelet coefficients are given to a 24-layer convolutional network as data (input) and labels (output). The inverse wavelet transform should be performed on the output results to find the normal-dose CT image. 

Generative adversarial networks (GAN) are a group of deep neural networks first introduced by Goodfellow \cite{gan}. GAN has two sub-networks, a generative network (G) and a discriminative network (D) that are trained simultaneously. The discriminative network is responsible for defferentiating real data from fake data while the generative network tries to create fake data as close as possible to the real data and fool the discriminator. Generative adversarial networks have attracted much interest, and researchers have applied it to different fields such as text to image synthesis \cite{gan-text-to-image}, image super-resolution \cite{gan-sr} and video generation \cite{gan-video}. GAN has also been used to remove noise from low-dose CT images \cite{ldct-gan-1,ldct-gan-2,sharpness-aware}, where the generative network receives the low-dose CT images. It generates normal-dose appearing images that the discriminative network cannot distinguish from real normal-dose images. 
\begin{figure*}[!tb]
      \centering
      
      {\includegraphics[width=\textwidth]{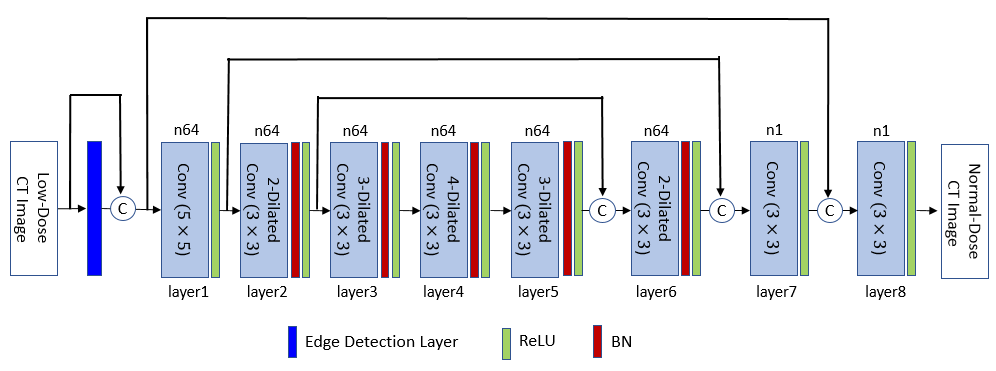}
      \caption{Architecture of the proposed network. BN stands for batch normalization, i-Dilated Conv represents convolution operator with dilatation rate i (i={2,3,4}) and the activation function is the rectified linear unit (ReLU). Operator \textcircled{c} performs concatenation.}
      \label{fig:proposed-network}
}

\end{figure*}
In this paper, we have proposed a deep neural network to remove noise from low-dose CT images. Figure \ref{fig:proposed-network} displays this network. One approach to achieving higher performance in deep learning is to increase the number of layers which has become possible after introducing residual learning \cite{residual-learning}, and batch normalization \cite{batch-normalization}. However, more layers essentially means more parameters and higher computational cost. In this research, we have looked for methods that enhance the efficiency of the network without adding to its complexity. For this purpose, our network employs batch normalization, residual learning and dilated convolution to perform denoising. We have also introduced an edge detection layer that improves the results with little increase in the number of training parameters. The edge detection layer extracts edges in four directions and helps to enhance the performance. Finally, we have shown that optimizing the mean-square error as the loss function do not capture all the texture details of the CT image. For this purpose, we have used a combination of perceptual loss and mean-square error (MSE) as an objective function that significantly improves the visual quality of the output and keeps the structural details. The perceptual loss is used in GAN to generate fake images that are visually close to the target image by comparing the feature maps of two images. Yang et al. \cite{ct-perceptual} have used the perceptual loss for CT image denoising but they compared the predicted image and the ground truth with one group of feature maps. In this study, feature maps have been extracted from four blocks of pre-trained VGG-16 \cite{vgg16} and used as a comparison tool in conjunction with the mean-square error.

\section{Methods}
\subsection{Low-Dose CT Simulation} \label{Low-Dose CT Simulation}

One of the challenges in applying machine learning techniques to the medical domain is the shortage of training samples. A neural network learns the probability distribution of the data from all the samples that it sees during the training process. If there are insufficient samples to train the network for all conditions, the predictions will not be accurate. To train a network for low-dose denoising, we generally need normal-dose and low-dose pairs. Obtaining such a dataset is not easy. For this reason, we have generated a simulated low-dose dataset from normal-dose CT images to be used for training besides two other datasets that we had.  

According to the literature, the dominant noise of a low-dose CT image in the projection domain has a Poisson distribution \cite{poisson-noise1,poisson-noise2}. Therefore, to simulate a low-dose CT image, we have added Poisson noise to the sinogram data of the normal-dose image. The following steps show this procedure \cite{simulate-low-dose-matlab,simulate-low-dose}:
\begin{enumerate}
\item Compute the Hounsfield unit numbers of the normal-dose CT image $HU_{nd}$ from its pixel values, using the equation \ref{eq:hu_conversion} \cite{HU-convert}(if the CT image has padding, it should be removed, first),
\begin{figure*}[!tb]
\centering
\begin{subfigure}{.24\textwidth}
  \includegraphics[width=0.97\linewidth]{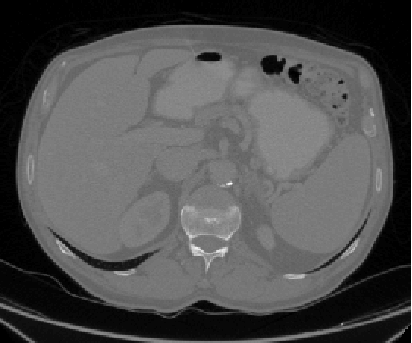}
  \caption{}
\end{subfigure}%
\begin{subfigure}{.24\textwidth}
  \includegraphics[width=0.97\linewidth]{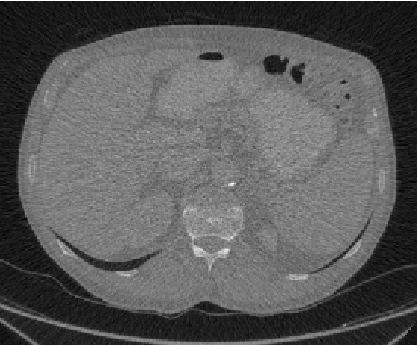}
   \caption{}

\end{subfigure}%
\begin{subfigure}{.24\textwidth}
  
  \includegraphics[width=0.97\linewidth]{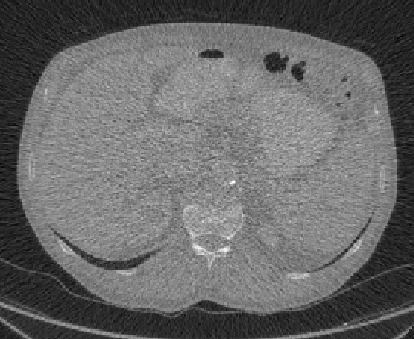}
    \caption{}

\end{subfigure}%
\begin{subfigure}{.24\textwidth}
  \includegraphics[width=0.97\linewidth]{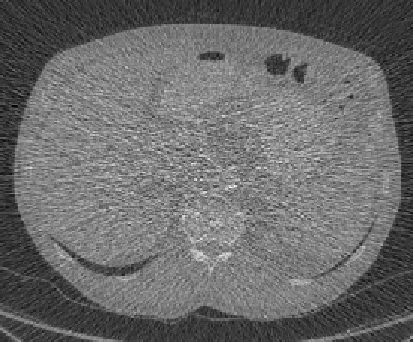}
    \caption{}
\label{fig:low-dose-simulation}
\end{subfigure}
\caption{Simulation of a low-dose CT image from an upper abdominal CT image, a) Normal-dose CT image, simulated low-dose image with b)$I^0_{ld}=1\times 10^4$, c)$I^0_{ld}=5\times 10^3$, d)$I^0_{ld}=2\times 10^3$}
\label{fig:simulated-low-dose}
\end{figure*}

\begin{equation}
\centering
\label{eq:hu_conversion}
HU_{nd} = \frac{Pixel Value}{Slope} + Intercept
\end{equation}
   
\item Compute the linear attenuation coefficients $\mu_{nd}$ based on water attenuation $\mu_{water}$,

\begin{equation}
\mu_{nd} = \frac{\mu_{water}}{1000}HU_{nd}+ \mu_{water}
\label{eq:HU-mu}
\end{equation}

\item Obtain the projection data for normal-dose image $\rho_{nd}$ by applying radon transform on linear attenuation coefficinets $\mu_{nd}$. To eliminate the size factor, this should be multiplied by the voxel size,
\item Compute the normal-dose transmission data $ T_{nd}$,
\begin{equation*}
\centering
T_{nd} = exp(\rho_{nd})
\end{equation*}    
\item Generate the low-dose transmission $ T_{ld}$ by injecting Poisson noise  \cite{simulate-low-dose},

\begin{equation}
T_{ld} = Poisson( I^o_{ld} T_{nd})
\label{eq:poisson}
\end{equation}
here, $I^0_{ld}$ is simulated low-dose scan incident flux.

\item Calculate the low-dose projection data $\rho_{ld}$,
\begin{equation*}
\label{eq:logarithm}
\rho_{ld} = ln(\frac{I^o_{ld}}{T_{ld}})
\end{equation*}
\item Find the projection of the added Poisson noise,
\begin{equation*}
\rho_{noise} = \rho_{nd} - \rho_{ld}
\end{equation*}

\item Compute the linear attenuation of the low-dose CT image $\mu_{ld}$,
\begin {equation*}
\mu_{ld} = \mu_{nd} + iradon (\frac{\rho_{noise}}{voxel})
\end{equation*}
where, $iradon$ represents the inverse Radon transform.
\item Finally apply the inverse of equation \ref{eq:HU-mu} to find the Hounsfield unit numbers for the low-dose CT image.
Figure \ref{fig:low-dose-simulation} demonstrates a normal-dose image and the simulated low-dose images with different incident flux $I^0$.
\end{enumerate}

\subsection{Dilated Convolution} 

Dilated convolution was introduced to deep learning in 2015 \cite{dilated,atrous} to increase the receptive field faster. Receptive field (RF) is the region of the input image that is used to calculate the output value. Larger receptive field means that more contextual information from the input image is captured. The classical methods to grow the receptive field is by employing pooling layers, larger filters and more layers in the network. A pooling layer performs downsampling and is a powerful technique to increase the receptive field. Although it is widely used in classification tasks, adoption of a pooling layer is not recommended in denoising or super-resolution tasks. Downsampling with a pooling layer may lead to the loss of some useful details that cannot be recovered completely by upsampling methods such as transposed convolution \cite{DAE_skip_connection}. Utilizing larger filters or more layers increases the number of parameters drastically, meaning larger memory resources will be needed. Dilated convolution, also called atrous convolution, can increase the receptive field with just a fraction of weights. One-dimensional dilated convolution is defined as
\begin{equation}
y[i] = \sum_{k=1}^{f}x[i+r.k]w[k]
\label{eq:dilated}
\end{equation}
where, $x[i]$ and $y[i]$ are the input and the output of the dilated convolution. $w$ represents the weight vector of the filter with length $f$, and $r$ is the dilation rate.\\

Receptive field of the layer $L$ ($RF_L$) with filter size $f\times f$ and dilation rate of $r$ can be computed from the equation \ref{eq:receptive_field} \cite{dilated-residual}.
\begin{equation}
RF_L=RF_{L-1}+(f-1)r
\label{eq:receptive_field}
\end{equation}

Equation \ref {eq:weights-number } computes the number of parameters needed for an N-layer convolutional network with a filter size $f\times f$.
\begin{equation}
number\, of\, weights = n\times f^2\times c+n^2\times f^2\times(N-2)+n\times f^2\times c
\label{eq:weights-number }
\end{equation}
here, $n$ is the number of filters in each layer and $c$ is the number of channels. For simplification, we assume all the layers have $n$ filters and the number of the channels in the input and output images are same. Table \ref{tab:filter-weights} compares the number of weights and layers needed to achieve receptive field equal to $13$ in different cases. 

\begin{table}[!tb]
\centering
\caption{Number of training weights to obtain $RF=13$ with different filter sizes. The number of filters in each layer is $n=64$}
\begin{tabular}
{|>{\centering}m{2.5cm}|>{\centering}m{1cm}|>{\centering}m{1cm}|>{\centering}m{1cm}|>{\centering}m{1cm}|}
\hline
\textbf{Filter size}& \textbf{$3\times3$ \newline$r=1$}&\textbf{$5\times5$ \newline$r=1$}&\textbf{$7\times7$ \newline$r=1$}&\textbf{$3\times3$ \newline$r=3$}\tabularnewline
\hline\hline
 {Number of layers \newline needed for RF=13} & 6 & 5& 4&4\tabularnewline
\hline
Number of weights & 148,608 &310400 & 407680 & 74880\tabularnewline
\hline
\end{tabular}

\label{tab:filter-weights}
\end{table}

To better understand the capability of dilated convolution, Wang et al. replaced the standard convolutions in \cite {gaussian_noise} with dilated convolutions with $r=2$ and achieved comparable performance in only 10 layers instead of 17 layers \cite{dilated-residual}. 

In this research, we have used an 8-layer dilated convolutional network to remove noise from low-dose CT images. The proposed network was inspired from a study by Zhang et al. \cite{prior}. The dilation rates used are 1, 2, 3, 4, 3, 2, 1, and 1 for layers 1 to 8. 

\subsection{Residual Learning}
One approach to improving the performance of a network is stacking more layers; nevertheless, researchers observed that networks with more layers do not always perform better. Contrary to  expectations, it has been seen that in a deeper network the training loss grows. This degradation problem implies that the optimization of a deep network is not as easy as a shallow one. He et al. \cite{residual-learning} proposed a residual learning framework to solve this problem by adding an identity mapping between the input and the output of a group of layers. The researchers have investigated many different combinations of adding shortcuts between different layers and achieved interesting results \cite{dense,DAE_skip_connection}.  

In this study, we have exploited the residual learning to improve the performance of the network. Our experiments showed that adding symmetric shortcuts between the bottom and top layers boosts the performance. As shown in Figure \ref{fig:proposed-network}, the input image and the output of layers 2 and 3 are concatenated with the output of layers 7, 6 and 5, respectively. These connections pass the details of the image to higher layers, as feature maps in the first layers contain more input information.

\subsection{Edge Detection Layer}
In image processing, edge detection refers to techniques that find the boundaries of objects in an image. Many of these methods search for discontinuities in the image brightness that are generally the result of the existence of an edge. Researchers have developed some advanced algorithms to extract edges from the image. In this study, we have adopted a simple edge detection technique to enhance the outcome of our network. Sobel edge detection operator \cite{sobel} computes the 2-D gradient of the image intensity and emphasizes regions with high spatial frequency by convolving the image with a $3 \times 3$ filter. The proposed edge detection layer is a convolutional layer that has four Sobel kernels as the non-trainable filters. The output edge maps are concatenated with the input image and given to the network. Our experiments confirm that the edge detection layer improves the performance of the network. 


\subsection{Objective Function}
Mean-square error (MSE) is widely used as an objective function in low-level image processing tasks such as image denoising or image enhancement. MSE computes the difference of intensity between the pixels of output and the ground-truth images. It is also used in many of the proposed algorithms for low-dose CT denoising. We started our research by optimizing MSE, but we noticed that the results do not express all the details of a CT image, despite peak signal to noise (PSNR) being relatively high. This problem has been seen in image super-resolution tasks too \cite{perceptual-loss}; however, it is more pronounced in Dicom CT images displayed with different grey-level mappings (windowing). Windowing helps to highlight the appearance of different structures and make a diagnosis. Our experiments showed that MSE loss generates blur images that do not include all textural details. 

Johnson et al. demonstrated that using a perceptual loss achieves visually appealing results \cite{perceptual-loss}. To compute the perceptual loss, the ground-truth image and the predicted image are given to a pre-trained convolutional network, one at a time. Comparison is then made between the feature maps generated by the two images. VGG-16 \cite{vgg16} is a pre-trained network for classification on ImageNet dataset \cite{imagenet} which is generally used to calculate the perceptual loss in generative adversarial networks.  

In this study, we have incorporated both MSE and perceptual loss to optimize the network. Our experiments showed that using the perceptual loss alone, results in a grid-like artifact in the output image. This effect has been perceived by other researchers, too \cite{perceptual-loss}. Therefore, we have combined both per-pixel loss $\mathcal {L}_{mse}$ and perceptual loss $\mathcal{L}_{P}$ to enhance optimization. 

\begin{equation}
\mathcal {L}(\theta) = \lambda_{mse} \mathcal {L}_{mse}(\theta)+ \lambda_P \mathcal{L}_{P}(\theta)
\label{eq:Loss}
\end{equation}
where, $\lambda_{mse}$ and $\lambda_P$ are weighting scalars for mean-squre arror loss and perceptual loss, respectively. The mean-square error between the ground-truth $y$ and the denoised image from the proposed network $\hat{y}$ is defined as,

\begin{equation*}
\mathcal{L}_{mse}(\theta)=||\hat{y}(\theta)-y||^2
\end{equation*} 

\begin{figure}
  \includegraphics[width=1\linewidth]{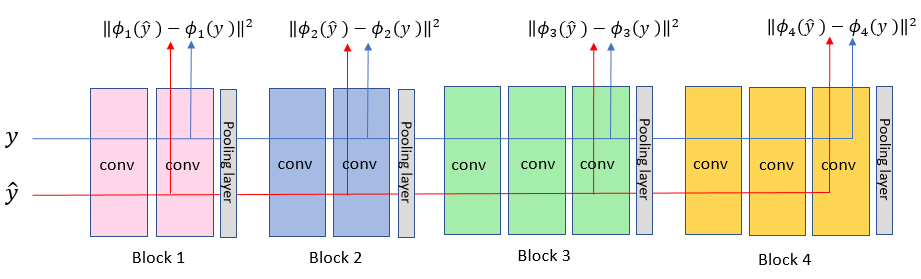}
  \caption{Perceptual loss is computed by extracting the feature maps of blocks 1, 2, 3, and 4 from a pre-trained VGG-16 network.}
    \label{fig:vgg16}
\end{figure}

Similar to other studies, we have employed VGG-16 network to measure the perceptual loss. In this study, we have extracted four groups of feature maps from VGG-16 network in different layers and used them to calculate the perceptual loss. As Figure \ref{fig:vgg16} demonstrates, we have used the output of the last convolutional layer (after ReLU activation and before pooling layer) in blocks 1, 2 , 3 and 4.  The perceptual loss function $\mathcal{L}_P(\theta)$ is
\begin{align*}
&\mathcal{L}_P(\theta) = \sum_{i=1}^{4}{\mathcal{L}_i(\theta)}\\
&\mathcal{L}_i(\theta)=\frac{1}{h_i w_i d_i}||\phi_i(\hat {y}(\theta))-\phi_i(y)||^2
\end{align*}
here, $\phi_i$ refers to the extracted feature maps from block $i$ with size $h_i \times w_i \times d_i$.  \\

Our experiments reveal that utilizing perceptual loss with the mean-square error greatly improves the visual characteristics of the output image.

\section{Experiments Setup} \label{Experiments Setup}
In this study, we have used three datasets to evaluate the performance of the proposed network in removing noise from low-dose CT images: simulated dataset, real piglet dataset, and a thoracic-abdominal (Thoracic) dataset. 

To create the simulated dataset, we downloaded lung CT scans \cite{lung-dataset} for a patient including $663$ slices from The Cancer Imaging Archive (TCIA) \cite{TCIA}. The CT images were taken with $330 mAs$ X-ray current tube, $120 KVp$ peak voltage and $1.25 mm$ slice thickness. Then, with the procedure explained in \ref{Low-Dose CT Simulation} we generated low-dose CT images. The incident flux of simulated low-dose CT ($I^0_{ld}$) in equation \ref{eq:poisson} is define as equal to $2\times 10^3$.  

The second dataset is a real dataset acquired from a deceased piglet. It contains 900 images with $100 KVp$,  $0.625 mm$ thickness. The X-ray currents for normal-dose and low-dose images are $300 mAs$ and $15 mAs$, respectively.  

Thoracic dataset \cite{thoracic-dataset} includes $407$ pairs of CT image from an anthropomorphic thoracic phantom. The current tube for normal-dose and low-dose CT images are $480 mAs$ and $60 mAs$, respectively with peak voltage of $120 KVp$ and slice thickness of $0.75 mm$.   

In each dataset, $70\%$ of the images are used for training the proposed network and $30\%$ for testing. Contrary to other studies that built a test dataset randomly, our test dataset holds the last $30\%$ of images in the original dataset. The reason is that the consecutive CT images are very similar to each other and testing the network on the random dataset does not clearly examine the effectiveness of the network on the new images. Using the last portion of CT images assures us that the testing is performed on images that the network has not seen before.  
To prepare the data for training the network, we have used pixel values of low-dose, normal-dose images divided by $4095$. This maps the data between $0$ and $1$ which is suitable for training neural networks.  

The original size of CT images in all the mentioned datasets is $512\times 512$. To boost the number of training samples, we have extracted overlapping patches of $40 \times 40$ from images, as the receptive field of the proposed network is $5+4+6+8+6+4+2+2= 37 $ in each direction. This also helps to reduce the memory resources needed during training. Since the network is fully convolutional, the input size does not have to be fixed, test images with their original size are fed to the network. To avoid boundary artifacts, zero padding in convolutional operators are used \cite{prior}. The activation function is rectified linear unit and the number of filters in all convolutional layers is $64$ except layers $7$ and $8$ which have $1$ filter. To see how adding the edge detection layer and utilizing MSE and perceptual loss improve performance, we have trained three networks. Training of all the networks are performed with Adam optimizer in two stages: with learning rate $1e-3$ for 20 epochs and then learning rate $1e-4$ for 20 epochs. Glorot normal is used to initialise the weights \cite{glorot}. The implementation was based on Keras with Tensorflow backend on system with an Intel core $i7$ CPU $3.4 GHz$, $32 G$ memory and GeForce GTX $1070$ Graphics Card.

\section{Results}
To evaluate the performance of the proposed network, we have compared the results with state of the art BM3D \cite{bm3d} and CNN200 network \cite{chen}. As mentioned earlier, the initial idea of the proposed network was derived from \cite{prior} designed for image super-resolution. To investigate how each change in the network architecture affects the performance, we have made three more comparisons with three more networks.   

The first network is designed to examine how residual learning enhances the outcome. This network is similar to the one in \cite{prior}, but there are shortcuts between the outputs of layers 2 and 3 with the outputs of layers 6 and 5, respectively. We call this network DRL (dilated residual learning). The objective function for this network is mean-square error (MSE), and we demonstrate that adding shortcut connections improves the results. 

In the second network, we have added the edge detection layer to the beginning of the network. This network is named DRL-E and is shown in Figure \ref{fig:proposed-network}. We have optimized this network by three objective functions to investigate the effects of choosing a loss function on the results. First, this network is optimized by MSE loss function and we refer to it as DRL-E-M. Next, we optimized the network by perceptual loss and it is called DRL-E-P in this paper. Finally, the proposed network is trained by the objective function defined in Equation \ref{eq:Loss}. This network optimizes a combination of mean-square error and perceptual loss to learn the weights and achieve the best results. We refer to this combination as DRL-E-MP. To verify that the improvements are the result of the network alterations, all the networks are trained with equal learning rates and epochs as explained in section \ref{Experiments Setup}.  

For each algorithm, we have provided the quantitative results proving that the proposed network DRL-E with dilated convolutions, shortcut connections, and the edge detection layer outperforms the other networks. Moreover,visual comparisons confirm that utilizing the proposed objective function improves the perceptual aspects of the DRL-E network further, and conserves most of the details in the image. 

\subsection{Denoising results on simulated lung dataset}
 
\begin{table*}[!t]
\caption{The average PSNR and SSIM of the different algorithms for the Lung dataset.}
\begin{center}
\begin{tabular}{|p{0.8cm}||p{0.8cm}|p{0.8cm}|p{0.8cm}|p{0.8cm}|p{0.8cm}|p{0.8cm}|p{0.8cm}|p{0.8cm}|}
\hline

\begin{turn}{-90}
Metric 
\end{turn}
&\begin{turn}{-90}Low-dose image   \end{turn}
&\begin{turn}{-90} BM3D \end{turn}
&\begin{turn}{-90}CNN200 \cite{chen} \end{turn}
&\begin{turn}{-90} \cite{prior} \end{turn}
&\begin{turn}{-90}DRL \cite{DRL}  \end{turn}
&\begin{turn}{-90} DRL-E-M  \end{turn}
&\begin{turn}{-90}DRL-E-P \end{turn}
&\begin{turn}{-90}DRL-E-MP\end{turn}
\\
\hline
PSNR & 14.59 & 24.76 &33.19 & 33.74 &34.17 & 36.64& 33.47&35.57 \\
\hline
SSIM &0.2008  &0.6750  & 0.8768& 0.8804 &0.9281 &0.9733 &0.5880 &0.6910  \\
\hline
\end{tabular}
\end{center}
\label{tab:psnr-lung}
\end{table*}

\begin{figure*}
\centering
\subcaptionbox{Low-Dose}{\includegraphics [width = 0.3\linewidth]{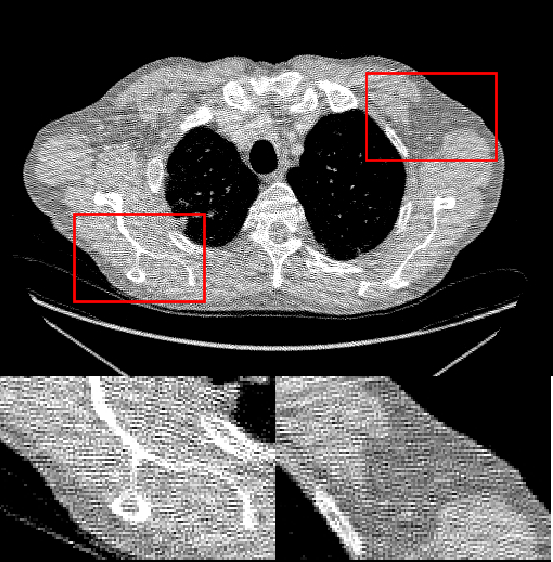}}\hspace{0.2cm}%
\subcaptionbox{Normal-Dose}{\includegraphics [width = 0.3\linewidth]{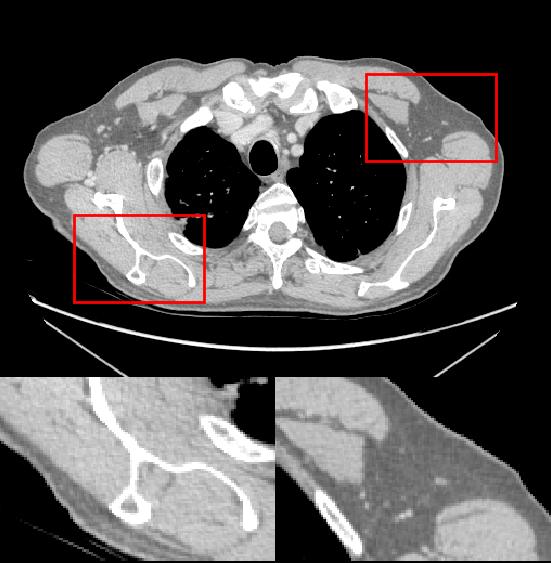}}\hspace{0.2cm}%
\subcaptionbox{BM3D}{\includegraphics [width = 0.3\linewidth]{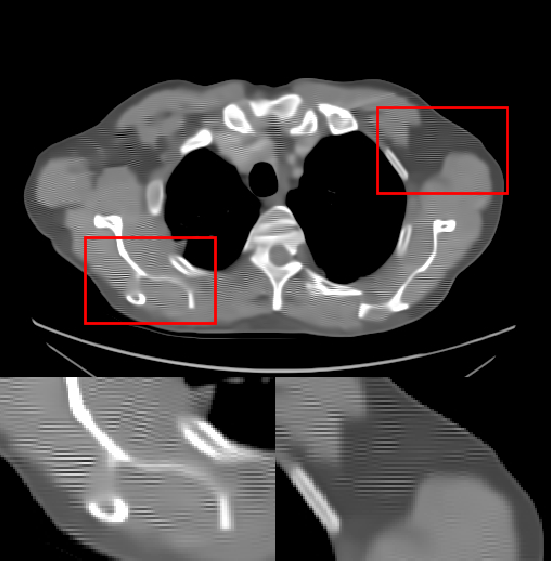}}\hspace{0.2cm}%
\subcaptionbox{CNN200 \cite{chen}}{\includegraphics [width = 0.3\linewidth]{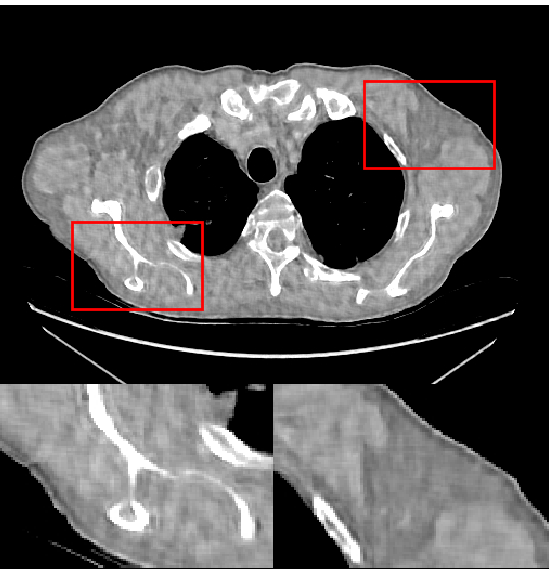}}\hspace{0.2cm}%
\subcaptionbox{\cite{prior}}{\includegraphics [width = 0.3\linewidth]{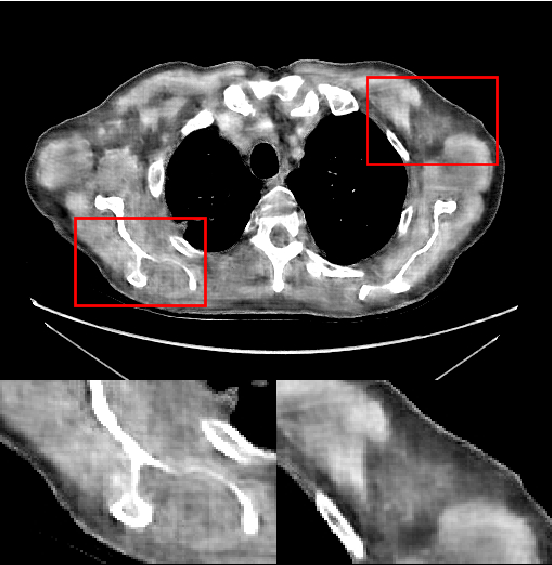}}\hspace{0.2cm}%
\subcaptionbox{DRL \cite{DRL}}{\includegraphics [width = 0.3\linewidth]{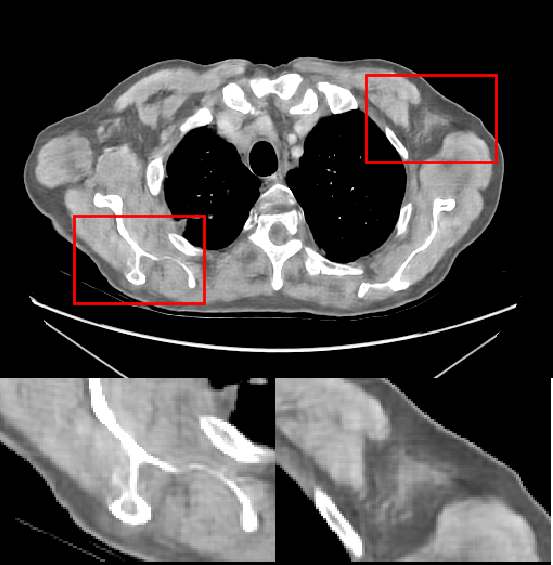}}\hspace{0.2cm}%
\subcaptionbox{DRL-E-M}{\includegraphics [width = 0.3\linewidth]{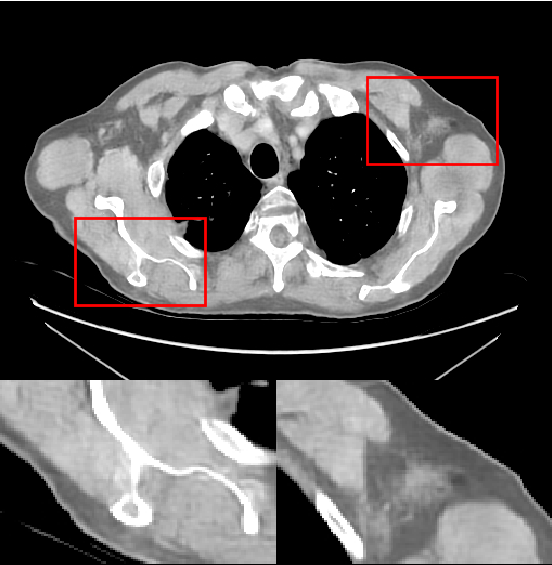}}\hspace{0.2cm}%
\subcaptionbox{DRL-E-P}{\includegraphics [width = 0.3\linewidth]{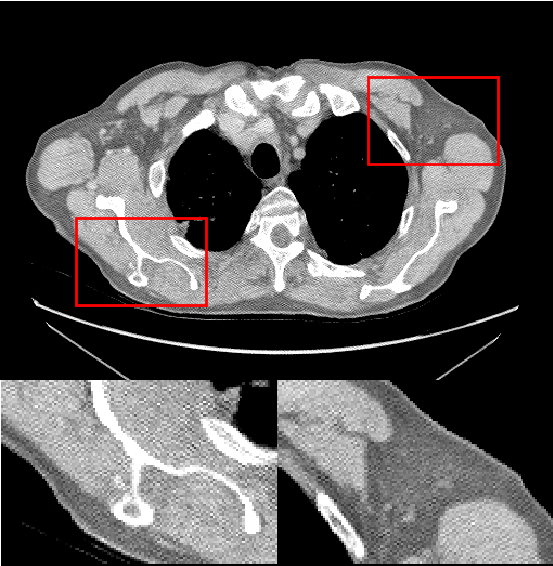}}\hspace{0.2cm}%
\subcaptionbox{DRL-E-MP}{\includegraphics [width = 0.3\linewidth]{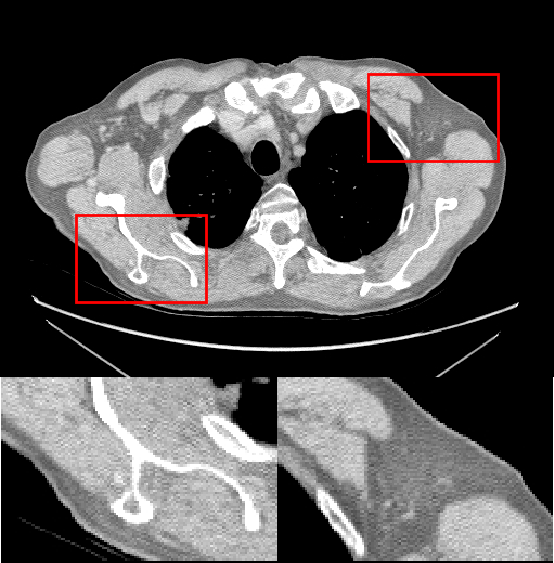}}

\caption{Denoising results of the different algorithms on the Lung dataset in abdomen window with select regions magnified.}
\label{fig:results-lung1}
\end{figure*}

\begin{figure*}
\centering
\subcaptionbox{Low-Dose}{\includegraphics [width = 0.3\linewidth]{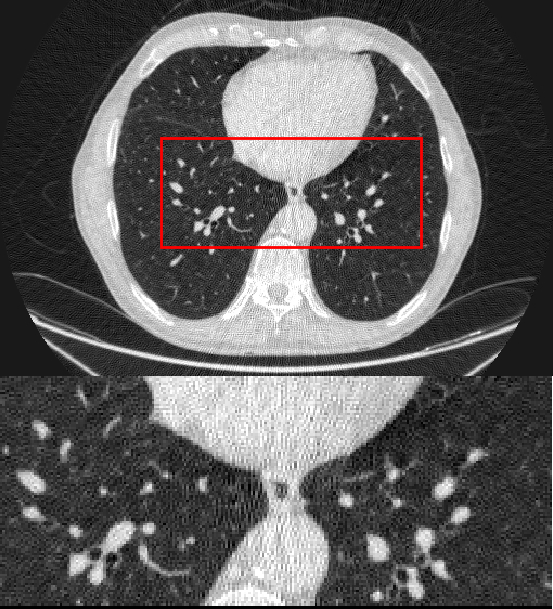}}\hspace{0.2cm}%
\subcaptionbox{Normal-Dose}{\includegraphics [width = 0.3\linewidth]{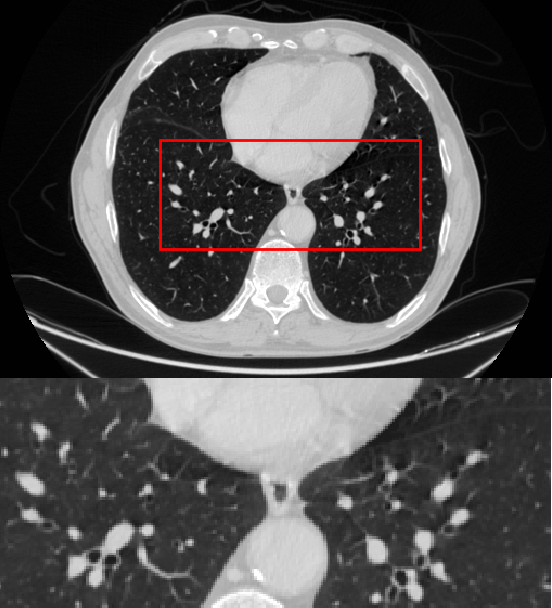}}\hspace{0.2cm}%
\subcaptionbox{BM3D}{\includegraphics [width = 0.3\linewidth]{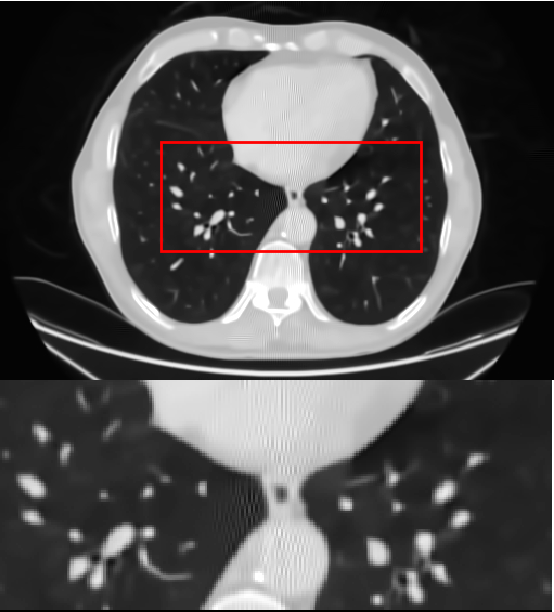}}\hspace{0.2cm}%
\subcaptionbox{CNN200 \cite{chen}}{\includegraphics [width = 0.3\linewidth]{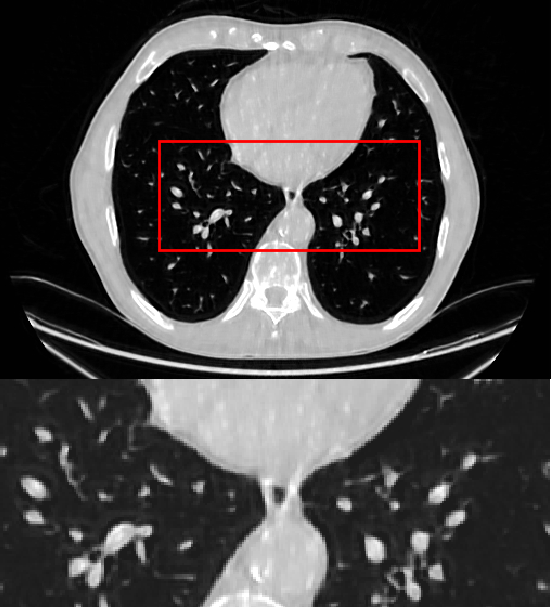}}\hspace{0.2cm}%
\subcaptionbox{\cite{prior}}{\includegraphics [width = 0.3\linewidth]{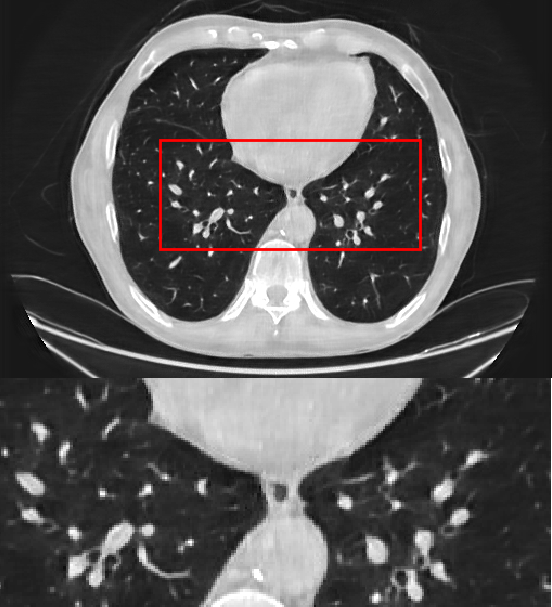}}\hspace{0.2cm}%
\subcaptionbox{DRL-E-MP}{\includegraphics [width = 0.3\linewidth]{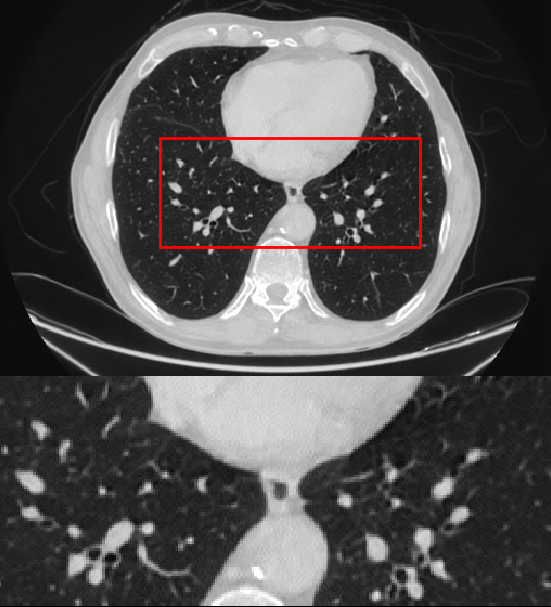}}

\caption{Denoising results of the different algorithms on Lung dataset in lung window.}
\label{fig:results-lung2}
\end{figure*}
Table \ref{tab:psnr-lung} displays the average peak signal to noise ratio (PSNR) and structural similarity (SSIM) of applying the state of the art BM3D and six neural networks. Figures \ref{fig:results-lung1} and \ref{fig:results-lung2} give the visual results for the Lung dataset in two different window. Windowing helps to visualize the details of CT images properly. Here, we have shown the results in the lung and abdominal window to distinguish the differences better.  Abdomen window helps to distinguish small changes in density and displays more texture details. Since lung is air-filled, it has very low density and appears black in the abdomen window. Lung window improves the visibility of the lung parenchyma  and areas of consolidation.  

Figures \ref{fig:results-lung1} and \ref{fig:results-lung2} demonstrate that the alterations of the network have enhanced the outcome step by step. This dataset demonstrates the effectiveness of perceptual loss. The results show using MSE as an objective function generates smooth regions and effects the details in the texture. On the other hand, perceptual loss forces the output of the network to be perceptually similar to the ground-truth. However, training the network solely by perceptual loss generates grid-like artifacts in the output image. As the results of DRL-E-MP demonstrate, the combined objective function saves most of the details in the textures and provides a better visual outcome.    

As one can expect, exploiting perceptual loss do not improve PSNR. The reason is that high PSNR is the result of low MSE. If a network is trained to minimize MSE, it will always have higher PSNR compare to a network that is trained to minimize the perceptual loss.  

\subsection{Denoising results on real Piglet dataset}
\begin{figure*}
\centering
\subcaptionbox{Low-Dose}{\includegraphics [width = 0.23\linewidth]{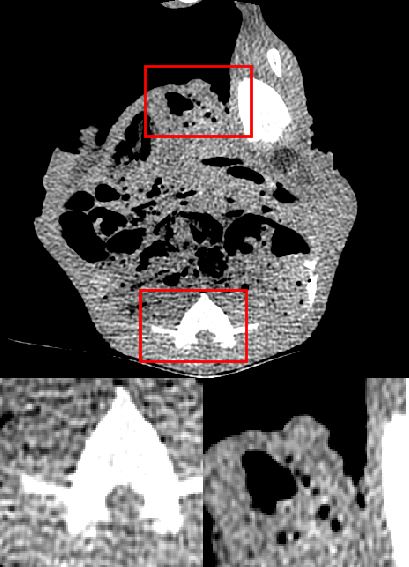}}\hspace{0.2cm}%
\subcaptionbox{Normal-Dose}{\includegraphics [width = 0.23\linewidth]{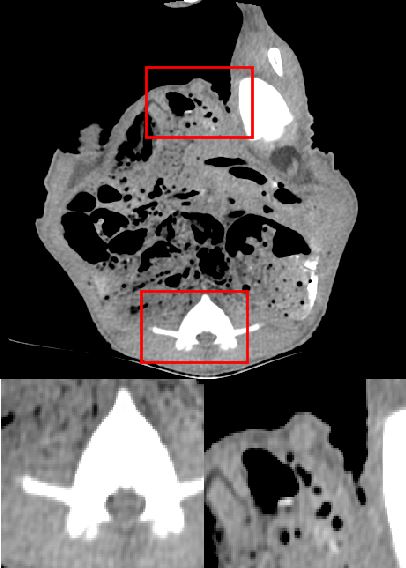}}\hspace{0.2cm}%
\subcaptionbox{BM3D}{\includegraphics[width = 0.23\linewidth]{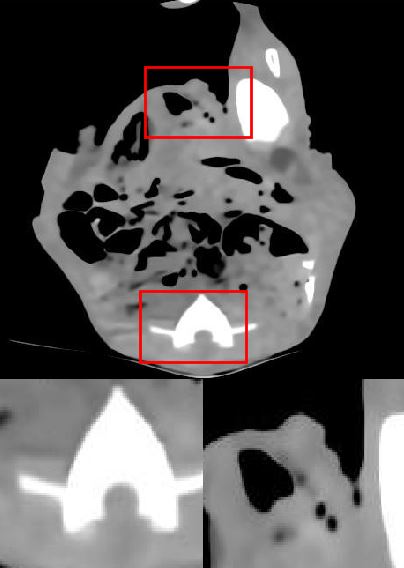}}\hspace{0.2cm}%
\subcaptionbox{CNN200 \cite{chen}}{\includegraphics [width = 0.23\linewidth]{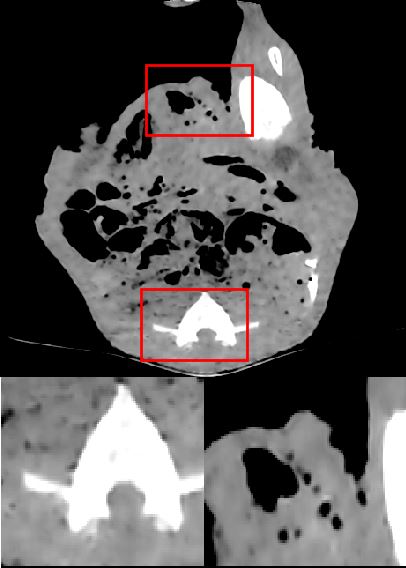}}\hspace{0.2cm}%
\subcaptionbox{\cite{prior}}{\includegraphics [width = 0.23\linewidth]{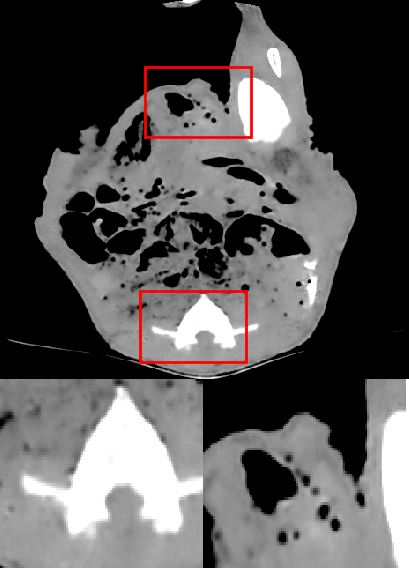}}\hspace{0.2cm}%
\subcaptionbox{DRL-E-M}{\includegraphics [width = 0.23\linewidth]{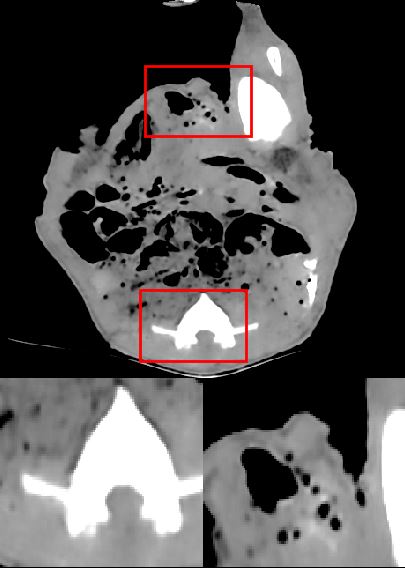}}\hspace{0.2cm}%
\subcaptionbox{DRL-E-P}{\includegraphics [width = 0.23\linewidth]{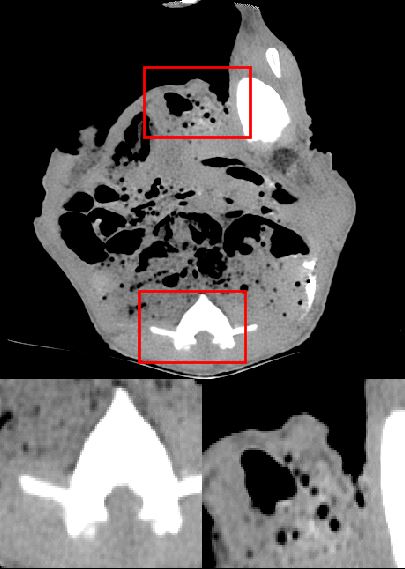}}\hspace{0.2cm}%
\subcaptionbox{DRL-E-MP}{\includegraphics [width = 0.23\linewidth]{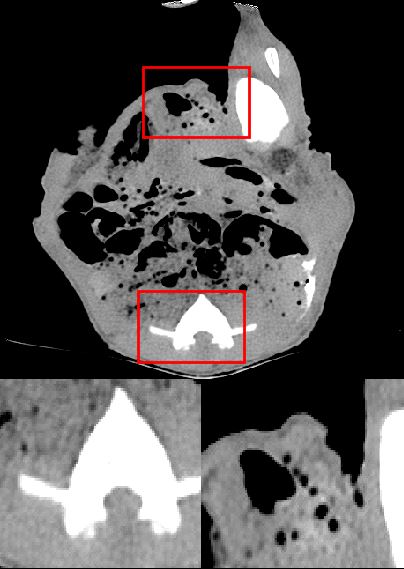}}

\caption{Denoising results of the different algorithms on an abdominal image from the Piglet dataset in abdomen window with select regions magnified.}
\label{fig:results-pig}
\end{figure*}

Table \ref{tab:psnr-pig} displays the quantitative effects of performing the denoising on real low-dose CT images for the Piglet dataset which approves the results obtained from the simulated Lung dataset. Comparing the PSNR coefficients of the BM3D,  CNN200, \cite{prior}, DRL \cite{DRL} and DRL-E demonstrates that when the objective function is MSE,  the network with residual learning and edge detection layer outperforms the other ones. Figure \ref{fig:results-pig} provides a visual comparison between the outcomes. It reveals that joining perceptual loss and per-pixel loss further improves the produced images by the proposed network. DRL-E-MP resembles the normal-dose CT image better by reconstructing the fine details.  

\begin{table*}[!t]
\caption{The average PSNR and SSIM of the different algorithms for the Piglet dataset.}

\begin{center}
\begin{tabular}{|p{0.8cm}||p{0.8cm}|p{0.8cm}|p{0.8cm}|p{0.8cm}|p{0.8cm}|p{0.8cm}|p{0.8cm}|p{0.8cm}|}
\hline

\begin{turn}{-90}
Metric 
\end{turn}
&\begin{turn}{-90}Low-dose image   \end{turn}
&\begin{turn}{-90} BM3D \end{turn}
&\begin{turn}{-90}CNN200 \cite{chen} \end{turn}
&\begin{turn}{-90} \cite{prior} \end{turn}
&\begin{turn}{-90}DRL   \end{turn}
&\begin{turn}{-90} DRL-E-M  \end{turn}
&\begin{turn}{-90}DRL-E-P \end{turn}
&\begin{turn}{-90}DRL-E-MP\end{turn}
\\
\hline
PSNR & 39.93& 41.46 &44.18 & 44.83 &44.96 & 45.10 &44.01&44.12 \\
\hline
SSIM &0.9705  &0.9733  &0.9804& 0.9816&0.9881 &0.9885 &0.9782 &0.9807  \\
\hline
\end{tabular}
\end{center}
\label{tab:psnr-pig}
\end{table*}

\subsection{Denoising results on phantom Thoracic dataset}
\begin{figure*}
\centering
\subcaptionbox{Low-Dose}{\includegraphics [width = 0.3\linewidth]{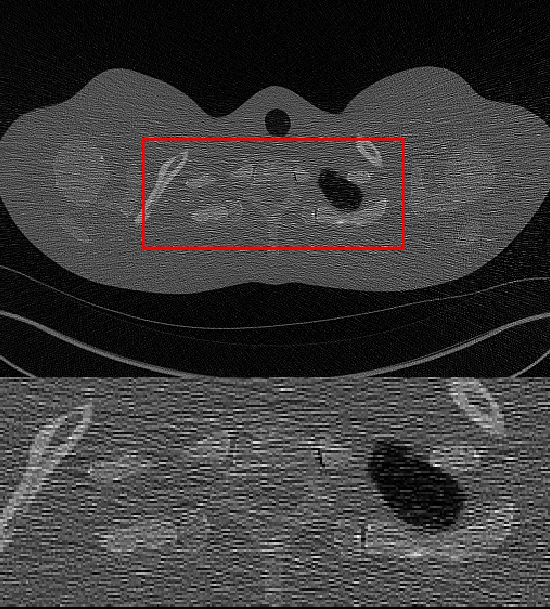}}\hspace{0.2cm}%
\subcaptionbox{Normal-Dose}{\includegraphics [width = 0.3\linewidth]{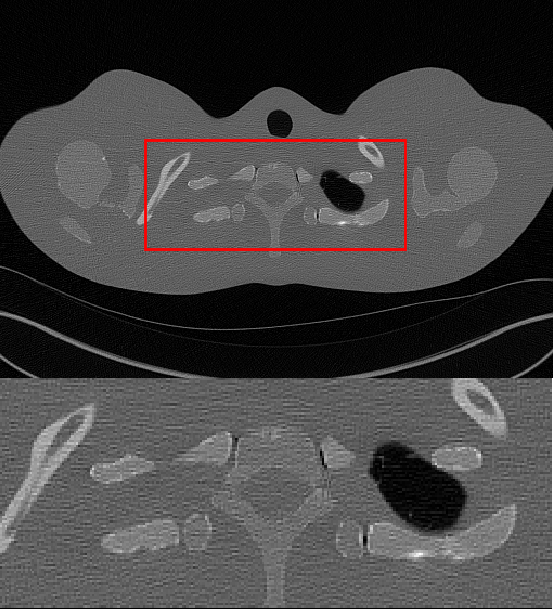}}\hspace{0.2cm}%
\subcaptionbox{BM3D}{\includegraphics[width = 0.3\linewidth]{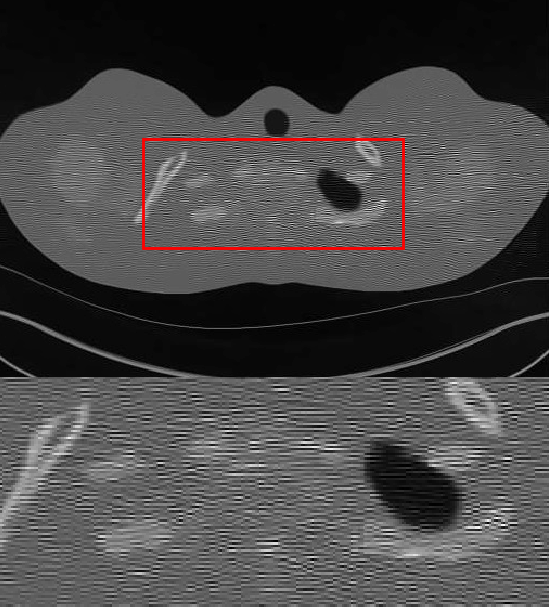}}\hspace{0.2cm}%
\subcaptionbox{CNN200 \cite{chen}}{\includegraphics [width = 0.3\linewidth]{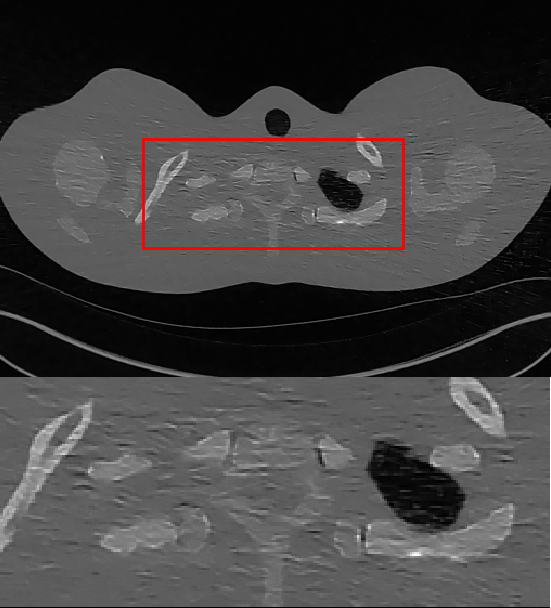}}\hspace{0.2cm}%
\subcaptionbox{\cite{prior}}{\includegraphics [width = 0.3\linewidth]{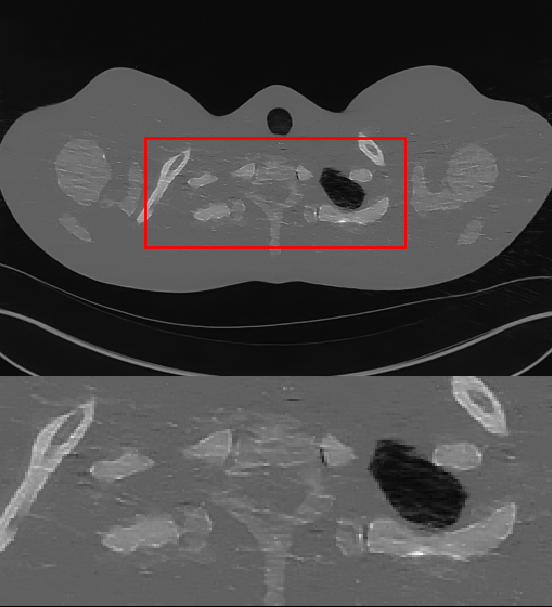}}\hspace{0.2cm}%
\subcaptionbox{DRL \cite{DRL}}{\includegraphics [width = 0.3\linewidth]{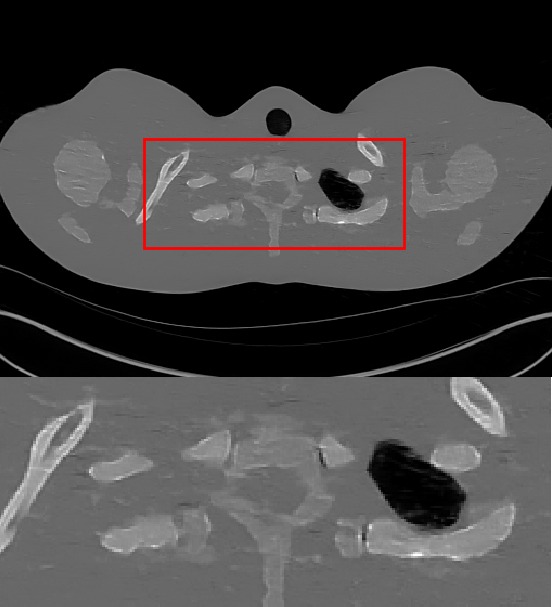}}\hspace{0.2cm}%
\subcaptionbox{DRL-E-M}{\includegraphics [width = 0.3\linewidth]{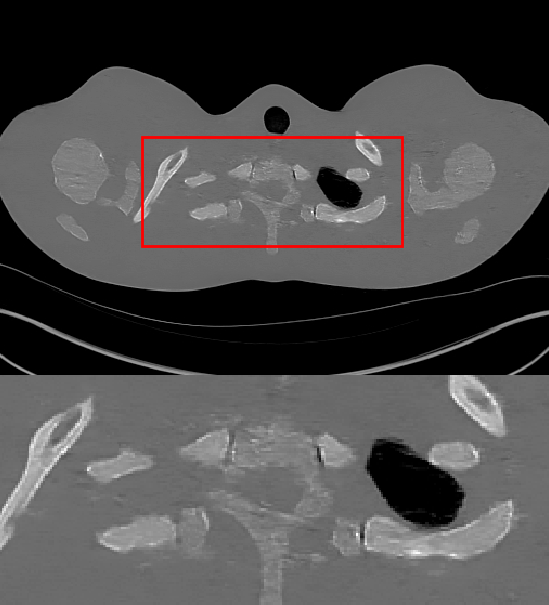}}\hspace{0.2cm}%
\subcaptionbox{DRL-E-P}{\includegraphics [width = 0.3\linewidth]{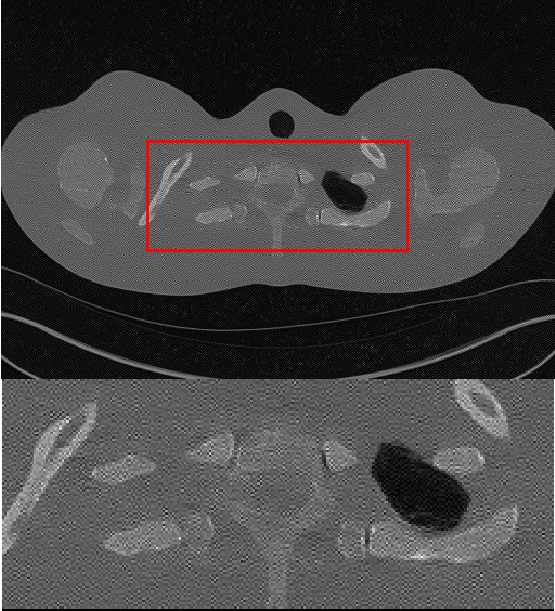}}\hspace{0.2cm}%
\subcaptionbox{DRL-E-MP}{\includegraphics [width = 0.3\linewidth]{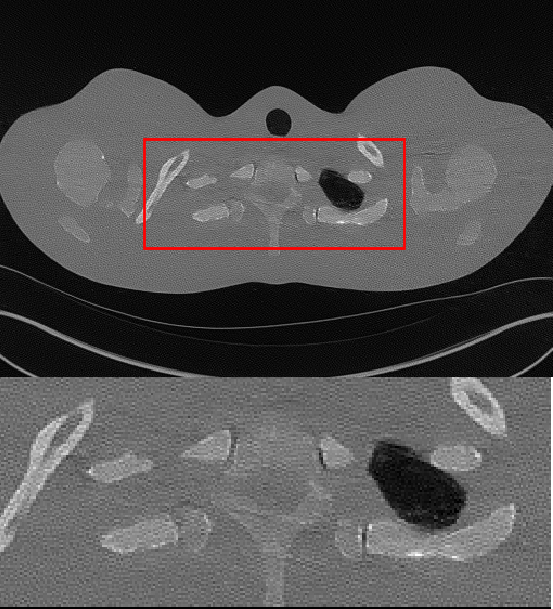}}

\caption{Denoising results of the different algorithms on Thoracic dataset in abdomen window.}
\label{fig:results-thoracic}
\end{figure*}

Table \ref{tab:psnr-thoracic} represents the PSNR and SSIM of denoising Thoracic dataset by all the methods. Results obtained for this dataset is consistent with the other experiments. Figure \ref{fig:results-thoracic} clearly exhibits the effects of each alteration. Comparing the results obtained by DRL and DRL-E-M confirms that the edge detection layer helps to deliver sharper and more precise edges. As explained before, the only difference between these two models is using the edge detection layer. 
\begin{table*}[!t]
\caption{The average PSNR and SSIM of the different algorithms for the Thoracic dataset.}

\begin{center}
\begin{tabular}{|p{0.8cm}||p{0.8cm}|p{0.8cm}|p{0.8cm}|p{0.8cm}|p{0.8cm}|p{0.8cm}|p{0.8cm}|p{0.8cm}|}
\hline
\begin{turn}{-90}
Metric 
\end{turn}
&\begin{turn}{-90}Low-dose image   \end{turn}
&\begin{turn}{-90} BM3D \end{turn}
&\begin{turn}{-90}CNN200 \cite{chen} \end{turn}
&\begin{turn}{-90} \cite{prior} \end{turn}
&\begin{turn}{-90}DRL   \end{turn}
&\begin{turn}{-90} DRL-E-M  \end{turn}
&\begin{turn}{-90}DRL-E-P \end{turn}
&\begin{turn}{-90}DRL-E-MP\end{turn}
\\
\hline
PSNR & 25.66& 30.86 &33.57 & 33.73 &34.02 & 34.03&26.25&31.50 \\
\hline
SSIM &0.4485  &0.6552  & 0.8001& 0.8018&0.8059 &0.8049 &0.4224 &0.6381  \\
\hline
\end{tabular}
\end{center}
\label{tab:psnr-thoracic}
\end{table*}

\section{Conclusion}
In this paper, we have combined the benefits of dilated convolution, residual learning, edge detection layer, and perceptual loss to design a noise removal deep network that produces normal-dose CT image from low-dose CT image. First, we have designed a network by the adoption of dilated convolution instead of standard convolution and also, using residual learning by adding symmetric shortcut connections. We have, also, implemented an edge detection layer that acts as a Sobel operator and helps to capture the boundaries in the image better. In the case of the objective function, we have observed that optimizing by a joint function of MSE loss and perceptual loss provides better visual results compared to each one alone. The obtained results do not suffer from over smoothing and loss of details that are the results of per-pixel optimization and the grid-like artifacts occurring with perceptual loss optimization. 

\begin{acknowledgements}
This work was supported in part by a research
grant from Natural Sciences and Engineering Research Council of Canada
(NSERC).
The authors would like to thank Dr. Paul Babyn and Troy Anderson for the acquisition of the piglet dataset.
The results shown here are in whole or part based upon data generated by the TCGA Research Network: http://cancergenome.nih.gov/.

\end{acknowledgements}

\bibliographystyle{spmpsci}      
\bibliography{reference}   

\end{document}